*Chapter 2*

# A Review of Assistive Technologies for Activities of Daily Living of Elderly

*Nirmalya Thakur* and *Chia Y. Han*

Department of Electrical Engineering and Computer Science,
College of Engineering and Applied Sciences,
University of Cincinnati, Ohio, US

## Abstract

One of the distinct features of this century has been the population of older adults which has been on a constant rise. Elderly people have several needs and requirements due to physical disabilities, cognitive issues, weakened memory and disorganized behavior, that they face with increasing age. The extent of these limitations also differs according to the varying diversities in elderly, which include age, gender, background, experience, skills, knowledge and so on. These varying needs and challenges with increasing age, limits abilities of older adults to perform Activities of Daily Living (ADLs) in an independent manner. To add to it, the shortage of caregivers creates a looming need for technology-based services for elderly people, to assist them in performing their daily routine tasks to sustain their independent living and active aging. To address these needs, this work consists of making three major contributions in this field. First, it provides a rather comprehensive review of assisted living technologies aimed at helping elderly people to perform ADLs. Second, the work discusses the challenges identified through this review, that currently exist in the context of implementation of assisted living services for elderly care in Smart Homes and Smart Cities. Finally, the work also outlines an approach for implementation, extension and integration of the existing works in this field for development of a much-needed framework that can provide personalized assistance and user-centered behavior interventions to elderly as per their varying and ever-changing needs.

* Corresponding Author's Email: thakurna@mail.uc.edu.

## INTRODUCTION

Quality of life can be enhanced by technology-based smart and intelligent assistant agents. Their effectiveness lies in their ability to address universal usability in diverse ways. Universal Usability [1] focuses on the three interrelated and interdependent areas of user diversity, technology diversity and bridging the gap between what users know and what they need to know [2]. User diversity refers to the differences in users in terms of their expertise level, age group, perceptual and cognitive differences, motor impairments, disabilities and knowledge. Technology diversity refers to a myriad of networked computer-based systems: from desktop computers, laptop computers, to portable devices, mobile phones with many screen sizes and connection speeds.

Speaking in terms of user diversity, the rapidly increasing population of elderly people has been one of the characteristic features of this modern century. At present there are around 962 million elderly people [3] across the world. This huge population of elderly people accounts for nearly 8.5 percent of the world's total population as shown in Figure 1.

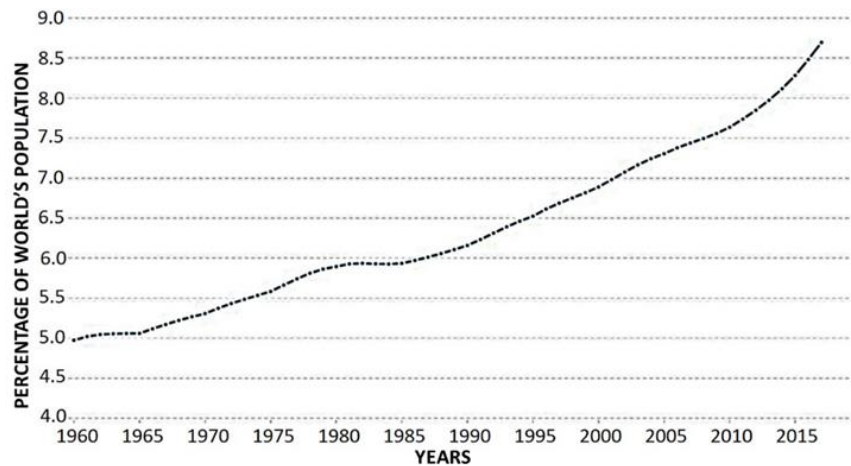

Figure 1. Elderly Population of the world expressed as a percentage of the total population.

Recent studies [4] have predicted that by the year 2050 the population of elderly people will become around 1.6 billion globally and will end up outnumbering the population of younger people worldwide. Their number is further expected to increase and reach 3.1 billion by the year 2100. With increasing age, the various needs in terms of personal, social and healthcare requirements increase.

"The United Nations Principles for Older Persons" [5] states a number of principles that may be adopted to enhance the quality of life experienced by elderly people. These principles focus on the necessity to design assistive environments that can adapt with respect to the dynamic needs of the elderly, to improve their physical as well as mental wellbeing. Increasing age is also associated with a number of health-related issues which poses an increased burden on the world economy. The number of elderly people across the world with dementia has doubled in recent times [6] and their number is predicted to again double by the year 2030, leading to approximately 76 million people with dementia worldwide.

In 2010 alone, approximately $604 billion costs were incurred to the healthcare industry in looking after people with dementia and this number is increasing at an alarming rate [6]. It is essential for modern day policy makers to develop systems and technologies that are equipped with

functionalities to adapt and address these needs of elderly people. The challenges for policy makers in this context, is to design technologies that are affordable and sustainable and can help to improve the quality of life experienced by elderly people. According to recent reports of the World Health Organization, the shortage of caregivers is a major concern in this regard. The current number of caregivers across the world is only 7.2 million and their population is expected to increase to around 12.9 million by 2035, which is significantly less as compared to the large scale predicted population increase of elderly people [7] as discussed above.

Aging has also been seen to be associated with declining patterns of user acceptance towards new technologies and experiences [8]. As per [9], 40% of elderly people face some form of limitations while performing physical activities especially in the context of performing their daily routine tasks [10]. A recent study [11] has indicated a total of eight typical problems associated with aging – (1) lack of self-sufficiency, (2) prone to accidents, (3) health issues, (4) neurological disorders, (5) loneliness, (6) psychological disorders, (7) financial status, and (8) heath care costs and management. Catering to all these needs of elderly as well as sustaining their independent living [12], is a challenge, especially in face of rapidly changing households as well as evolving technology based living spaces [13] for example – Smart Homes and Smart Cities. In a broad sense, a Smart Home can be defined as a technology-based living space that provides homeowners with assistance, safety, luxury, fosters their overall well-being and contributes towards an enhanced living experience through interconnected devices and systems that operate together through sharing user data and automating actions, through an Internet-of-Things (IoT)-based environment [7].

The essence of providing technology-based solutions to address these challenges lies in the effectiveness of technology to address the diversity in elderly population which can be broadly characterized by their varying age group, gender and differences in their background [14]. This diversity in the elderly population leads to varying experiences resulting in different habits and diverse nature of user interactions. Elderly people can broadly be subdivided into two sub-groups based on their age – (1) Young elderly – aged 65 to 85 and (2) Old elderly - aged 85 and above.

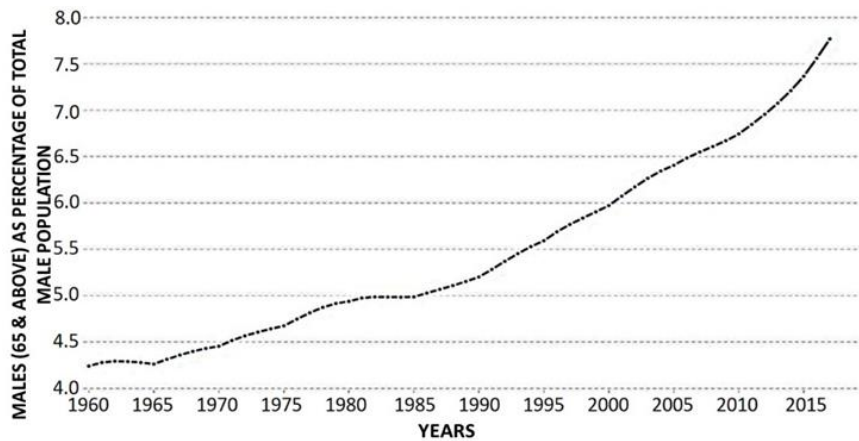

Figure 2. Males (65 and above) expressed as a percentage of the total male population.

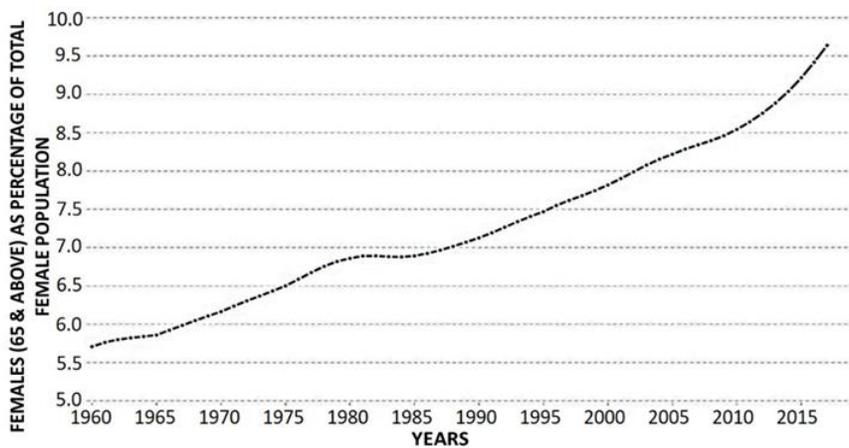

Figure 3. Females (65 and above) expressed as a percentage of the total female population.

These specific sub-groups are also diverse in terms of their gender and recent research [6, 7] has shown that this also leads to different characteristics of user interactions in elderly people. Studies on world population data [3] has shown that females aged more than 65 years account for around 10% of the worlds total female population while males aged more than 65 years account for just 8% of the world total male population. This is shown in Figure 2 and Figure 3.

Researches in this field have investigated the impact of a single or several challenges in the context of assistive technologies; however, there is a need for a study that reviews all the existing works in this field and takes a comprehensive approach towards analyzing the various challenges and limitations that assistive technologies face in the context of their implementation in living spaces for addressing the varying needs of elderly with respect to their diversities. Such a study would help policy makers and governments to work towards improving the quality of lives of elderly people and meeting their varying needs. This forms the main motivation for this work.

The concept of Assistive Technology can be broadly defined as systems and resources that take a holistic approach towards tracking, monitoring and fostering the overall health, security, safety and quality of life of its users. It can include systems, softwares, products, devices, practices as well as a combination of any two or more of these to address the needs of its users who have some form of limitations or disabilities [7]. In the context of Smart Homes, the application of assistive technologies in the future holds immense potential for fostering independent living and active aging for elderly people.

The work is organized as follows: Chapter 2 discusses about Activity Theory and Activities of Daily Living. Chapter 3 presents a review of recent advances in the field of Assistive Techbologies with a specific focus on their implementation of in the context of ADLs. This is followed by a discussion in Chapter 4 about the challenges that exist in the context of implementation of assistive technologies for healthy and independent aging in Smart Homes. The conclusion and scope for future work is presented next, which is followed by references.

## ACTIVITY THEORY AND ACTIVITIES OF DAILY LIVING

### Activity Theory

In a broad manner, activity may be defined as a specific relationship between a subject (an actor) and an object (an entity existing in the world).

Activity is commonly represented as "S <-> O" which represents the relationship and interaction between a subject and an object [15]. This relationship between the subject and object is governed by two specific features – (a) subjects act on objects because they have specific needs which they seek to accomplish and (b) subjects and objects mutually determine each other in a given context.

(a) Subjects and their associated needs: Activities are considered as a way of life [15] in this objective world, which is associated with different needs at different phases. To meet these needs, a person needs to interact with objects.

(b) subjects and objects mutually determine each other: The actions of the subject on any object in a given world, depends on the attributes of the object as well as the nature by which those attributes influence the subject. For instance, if a person has to solve a mathematics problem, the person needs to have skills in mathematics. Similarly, the difficulty of the problem would determine the extent to which the person would use his or her skills to solve the given problem.

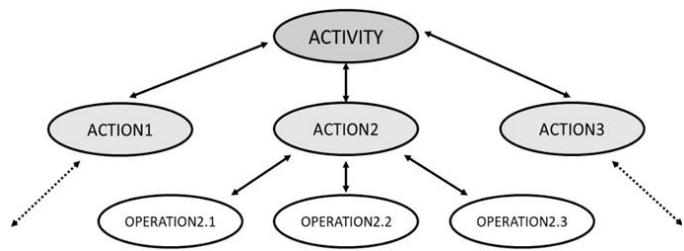

Figure 4. Activity expressed in terms of action-operation model.

In addition to these relationships between a subject and an object that determine the activity, there is also another characteristic feature which determines an activity, it is called motive. Activity theory [15] defines motive as the characteristic of an object to meet a certain need of the subject. At a closer level, it is observed that this interaction between a subject and an object does not necessarily take place in a linear manner, instead, it takes place in a hierarchical manner with needs and motives associated to objects at each level of the hierarchy. For instance, let us consider a person typing something in a specific application in a computer screen. The specific letters typed by the person are associated with the need to form meaningful words which are associated with the need to form meaningful sentences. This is further associated with the need to form collection of sentences or a paragraph and this continues. These aspects of the activity constitute different levels of the hierarchy which needs to be accomplished. Therefore, it may be concluded that needs are the governing factors behind activities.

Needs may be viewed from two different perspectives – biological or psychological. From a biological perspective, needs are associated with specific objects in the environment that causes the subject to act on them and accomplish a desired task through the activity. From the psychological perspective, needs are not associated with objects but are associated with specific states of the human body (for instance the feeling of hunger) which causes the subject to look for a specific object to act upon and satisfy this need. Activities are also not monolithic in nature. According to the initial concept of activity theory [15], an activity is a hierarchical structure which consists of three levels in the hierarchy as shown in Figure 4.

The uppermost layer is the activity that is associated with a specific motive that causes the subject to act on a specific object. The total process of the activity occurs in a step by step manner with each step leading to the motive. Therefore, it is understood that each of these steps may not directly be related to the motive but indirectly lead to it through the sequence or chain of steps. Activity theory defines these steps as actions and the respective objects at which these actions are performed are called goals. Accomplishment of goals associated with a given activity leads to the task of successfully completing the activity. In fact, actions can also be sub divided into a lower unit in the hierarchy – called operations. Operations are mostly characterized as routine processes that lead to the subject adjusting a specific action as per the given context. For instance, the action of walking might need to be adjusted differently in a side walk which is crowded, to avoid any collisions. However, recently the work based on Activity Centric Computing at Apple [16] has introduced another level called tasks in this hierarchical structure to represent an activity.

There are many important aspects in the activity theory, the following five are the ones that we consider key principles of activity theory:

(a) Object-Orientedness: This principle relates to the fact that activities are associated with objects and activities may be differentiated from each other based on different objects that are associated with them. Different needs of different subjects cause them to act on objects in their environment, specific to these needs and as per the varying motives of the subject.

(b) Hierarchical Nature: The initial concept of activity theory [15] defines activity as units of life which are represented by specific relationships between subjects and objects following a hierarchical manner of interaction as shown in Figure 4.

(c) Mediation: It is this concept of mediation in activity theory that differentiates the nature by which humans and animals perform activities. Mediation, in the context of activity theory, is mostly popular in humans. It refers to the structures and tools (for example – spoons, clothes etc.) that act as mediating objects facilitating the interaction between subjects and objects as per the needs of the subject.

(d) Internalization and Externalization: This explains that each activity is composed of both internal components as well as external components, the presence of one or both may or may not be evident based on the subject, specific needs, context and the world in which the subject is present. For instance, when children learn to count, they sometimes use their fingers for help but with time they no longer require doing the same and develop the ability to perform the counting without using fingers. Speaking of externalization, it is one of the important characteristics of human activity. For instance, a pen allows a poet to write down his or her thoughts, which leads to the activity of writing a poem.

(e) Development: Interactions of subjects on same or similar objects change with time. Activity Theory states that the nature of transformation of these interactions need to be studied to analyze these changes to understand varying nature of activities and their associated contexts at an intricate level. Activity Theory focuses on the importance of studying these transformational patterns in activities over time for researchers however it does not explicitly mention any specific method of study as different activities in different contexts might require different methods of study.

Recent research [15] in the field of Activity Centric Computing, has added another dimension to the concept of activity which was previously broadly defined as an interaction between subject and object. This had led to the addition of another element called 'community' to the subject-object model, which resulted in a model that facilitates three-way interaction as shown in Figure 5, listed below. This work also underlines the specific nature of interactions between these components as discussed next.

(a) Structures or tools are used for the subject-object interactions
(b) Specific set of rules are used for subject-community interactions
(c) Division of labor is used for community-object interaction

There can also be community-community interactions depending on the context, specific needs, motives and nature of interactions performed. For instance, consider a user experience designer who is working in a design team to design an application for forecasting weather conditions. In addition to working on designing the application, which in this case is the subject-object nature of interaction, this employee needs to interact with other members of his or her community, which in this case refers to the design team. These interactions may be in the form of group meetings, progress updates and similar activities. Next, the design team as a community also needs to communicate with other departments of the company by conveying updates about their work and discussing further plans. This refers to the community-community interaction for this example.

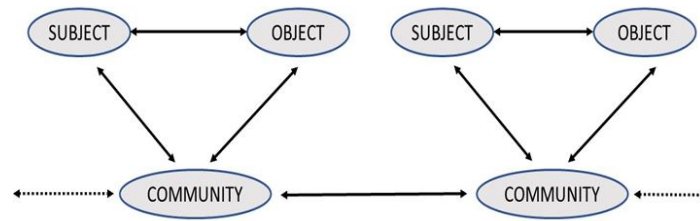

Figure 5. The subject-object-community interaction in Activity Theory.

## Activities of Daily Living (ADL)

Activities of Daily Living or in short ADLs refers to the daily routine activities and self-care activities which are mostly in the context of the premises of one's home. There are multiple definitions to the exact list of activities that comprise ADLs but in a generic manner, ADLs can be classified into the following five categories: Personal Hygiene, Dressing, Eating, Maintaining Continence and Mobility.

(a) Personal Hygiene: Refers to activities like showering, grooming etc.
(b) Dressing: Refers to activities that involve changing clothes, putting on new clothes and similar tasks
(c) Eating: This refers to eating different meals during the course of the day and as per ones eating habits.
(d) Maintaining continence: This primarily comprises the activity of using a restroom
(e) Mobility: This involves the activities of moving around from one place to another as per the requirement.

The ability of an individual to carry out these activities on their own or to seek help from caregivers gives an indication of the level of independence of the person [17]. Elderly people due to their increasing age which comes with associated disabilities, are quite often unable to accomplish these tasks which increases the need and relevance to provide technology-based solutions for their aid.

**Activities of Daily Living Characteristics**

Based on the varying needs and motives behind a user interacting with an object or certain objects in the given context, there may be the following general ways in which activities may be broadly performed [18]:

(a) Sequential Activity: This refers to the user doing one activity at a time. When the user finishes off with the given activity at hand, the user initiates with the next activity.
(b) Concurrent Activity: This refers to the situation when the user would be doing more than one activity at a given time instance. Although these activities may or may not start or end at the same time.
(c) Interleaved Activity: Real life activities are mostly interleaved. This refers to the user starting off an activity in the middle of another activity and returning to the former after some time which may or may not be after completing the new activity that the user started.
(d) False Start: This refers to a situation like an Interleaved Activity, but the only difference is the user does not return to the initial activity to complete it. This may be because the other activity required more attention or could also be that the initial activity was no longer relevant.
(e) Social Interactions: This refers to activities that are performed by individuals in a social context. Such activities involve more than one person working towards the same goal with similar needs and motives.

## REVIEW OF ASSISTIVE TECHNOLOGIES IN ADLS

The essence of designing assisted living technologies for catering to the needs and requirements of elderly during ADLs involes accurate activity recognition. To conduct this review, we conducted an extensive search across disciplines and sources including but not limited to SCOPUS, Web of Science and Google Scholar, by using the search terms, "assistive technologies," "assisted living," "elderly people." We were able to identify more than a 100 papers that met these search categories. As the specific focus of this review is to discuss the Assistive Technologies in the context of ADLs, so we identified only those works by reviewing the literatures and the citations of each. Based on this, we were able to develop a list of 32 works which are being reviewed next.

There are broadly two major approaches for recognition of ADLs which are (i) Sensor-Based Recognition and (ii) Vision-Based Recognition [18].

Sensor Based Recognition of ADLs: With the rapid advancement of both wearable and wireless sensor systems, these technologies have provided solutions for activity recognition. These sensors collect various kinds of data and then machine learning and pattern recognition principles are generally used to infer about the activity from the obtained data. Wearable sensors provide measurements from units like accelerometers, gyroscopes, heart rates, skin temperatures, galvanic skin response and similar entities, whose data are analyzed to infer information about the activity being performed. Wireless sensors mostly provide information about the gait orientation which is then analyzed by learning models to infer about the activity which was performed by the user. However, there are two major limitations of this approach. First, real world movements are generally complex and involve social interactions, so it is quite often difficult to analyze such movement patterns. Second, wearable sensors have issues like battery life which come in the way of experiments. Moreover, the user acceptance has been a problem; people are often hesitant to use wearable sensors without having the full knowledge of the same.

Vision Based Recognition of ADLs: This approach mostly involves using cameras and computer vision principles to analyze human behavior and understand the ADLs. Different types of cameras have been used by researchers which have different functionalities that allow researches to interpret the activity. The obtained data from these types of devices is mostly in the form of frames separated from real time videos or continuous video data which contains information about body motion features, gait orientation and similar. This data is then analyzed by computer vision principles to infer details about the behavior of the user in the same.

The way these ADL recognition models work can be classified into two categories – knowledge-based models and data driven models. Knowledge based approaches mostly make use of ontologies to infer about activities. Azkune et al. [19] developed a model which comprised of multiple layers to examine human actions in the context of daily routine tasks. The structure consisted of functionalities that were related to acquisition of information gathered from sensors' that tracked user actions, understanding those semantics, and developing an understanding of activities. In Riboni et al.'s [20] work, ontologies weren't utilized to infer the particular action or task being done by the user but they were utilized to directly confirm the activity based on the small tasks and actions. They created a knowledge base of activities which included tasks and actions performed on environment variables with reasoning principles to infer activities from small actions or tasks. Nevatia et al. [21] developed a language, based on image recognition and video data analysis inferring activities in real time.

Data driven approaches for activity recognition consist mostly of various data science and machine learning methods for study of human activities. In [22], Kasteren et al. proposed use of Hidden Markov Models (HMM) to study and analyze activities. Each state of the HMM was assigned a specific task in the context of completing the overall function of recognizing user behavior and inferring the activity being performed. The model not only recognised the user actions but also tracked the rate at which these actions changed to better infer the sequence of actions related to the given activity being performed. Cheng et al. [23] proposed a multilayered learning model, based on kernel techniques to infer user actions and recognize behavior in the context of activities. They also extended this model to perform activity analysis of multiple users in the given Internt of Things (IoT)-based environment. Skocir et al. [24] developed a framework that used sensor data and Artificial Neural Networks for performing activity recognition. They tested the system on studying enter and exit events through a door in a Smart Home environment and their system achieved high levels of performance. Doryab et al. [25] used machine learning methods to design an action recommender system that would recommend actions and tasks, based on the sequence of other tasks being performed upto that point, to medical practitioners in a typical IoT-based hospital environment.

Abascal et al. [26] proposed an indoor navigation system that used multiple sensors in an IoT environment to track the users current position and help them navigate to a specific region of the environment, based on the activity they wanted to perform, through means of an intelligent wheelchair aimed to help elderly people with disabilities easily move around. They also developed a handheld device that would be useful for elderly people with cognitive impairments to better understand their surroundings and interact with the same to perform their activities of interest. An activity monitoring system was developed by Chan et al. [27], specifically for studying elderly behavior in Smart Homes. In addition to studying human behavior, the system also involved an alert system to alert caregivers during a fall. In the work [28], Yared et al. proposed a multifunctional framework which possessed functionalities to first, analyze user actions in a kitchen environnment and second, infer any abnormalities in user behavior and environment parameters to automatically detect accidents.

Kim et al. [29] developed a system to track the overall well being of elderly people in IoT-based living spaces. Their system had a database consisting of the amount of time a user usually spends in the context of interacting with various environment parameters and they developed a learning model which could detect anomalies in terms of a user spending more or less time during their user interactions with these environment parameters. The objective of this work was to study the mental well being with a specific focus on analyzing user actions and activities. In the work [30] by Deen, a host of wireless sensors and wearables were used to develop a framework that could track, record and analyze mutimodal aspects of physiological signals from elderly behavior and send alert signals during the event of anomalies in one or more readings. Civitarese et al. [31] proposed an intelligent system that could analyze multimodal aspects of user interactions to infer about elderly people successfully completing Activities of Daily Living (ADLs). The system was able to analyze user interactions in this context to understand about symptoms of Mild Cognitive Impairment (MCI) in elderly people.

Iglesias et al. [32] developed a system which could analyze the touch-based interactions of users and infer their health status. The framework also included an alert system which would alert caregivers when the users health status revealed that they needed help and attention. A similar work was done by Angelini et al. in [33], where they developed a smart bracelet to track the health status of the user which also had the alerting capabilities. As an additional feature, the bracelet could also remind users of medications and other scheduled tasks during the day, if the user forgot the same.

A socially assistive robot was developed by Khosla et al. [34] for assisting elderly people perform ADLs in Smart Home environments. They also conducted usability studies to analyze user acceptance of this robot by elderly people. An intelligent decision making framework for augmenting user performance in the context of ADLs was developed by Tarik et al. [35]. It could not only study and track user behavior but it could also recommend actions to users if they needed help completing the given activity at hand. Sarkar [36] proposed an intelligent robot called "NurseBot" that consisted of a scheduler to maintain the information about different medications that need to be taken by elderly people.

The recent works of Thakur et al. [37-50] have extensively contributed to the field of proposing various assistive technologies for elderly care in Smart Homes. In [50], the authors proposed a Complex Activity Based Emotion Recognition Algorithm (CABERA), for analyzing the emotional state of the user in the context of ADLs. Their work was able to classify the emotional state of the user into six basic emotional states – happiness, sadness, anger, frustration, disgust and anxiety. In [46-48], they proposed a framework which could analyze the user experience and affective state of the user in the context of ADLs in a Smart Home environment. In [49], they proposed an approach to analyze the user acceptance of virtual assistants in the context of interaction between elderly people and virtual assistants to help the former perform ADLs. In [42, 43] the authors proposed approaches that can analyze the multimodal aspects of human behavior in the context of ADLs to understand user performance and provide timely assistance based on need. As an extension of their works, they proposed an improved activity recognition approach in [40], that outperforms most of the other activity recognition approaches in this field and can be readily implemented in Smart Homes. In [41], they developed a system, that can track and analyze user interactions in the context of ADLs and provide task recommendations based on context attributes for helping elderly people perform ADLs in an independent manner.

## CHALLENGES FOR ASSISTIVE TECHNOLOGIES

The field of assistive technologies holds immense potential for addressing the varying needs and requirements of the constantly increasing elderly population and contributing towards their independent living and healthy aging in the future of Smart Homes and Smart Cities. However, there remains several challenges, many of them are psychological and sociological factors beyond technical ones, for large scale implementation of such technologies for elderly care. Some of these as mentioned in [51] are (i) security and privacy of user data, (ii) building trust on technologies, (iii) ease of use, (iv) affordability, (v) suitable training and guidance for adoption, (vi) user acceptance, (vii) perception of no need, (viii) fear of dependence, (ix) feelings of embarrassment, (x) loss of dignity, and (xi) lack of accessibility and social inclusion.

Thus, in the context of development and implementation of assistive technologies, it is important to have a holistic view and to address these limitations for ensuring seamless adoption of such technologies for contributing towards Ambient Assisted Living of the global elderly population in the future of Smart Homes and Smart Cities.

## CONCLUSION AND FUTURE WORK

This work presents a rather comprehensive review of Assistive Technologies for ADLs of elderly people. Based on the reviewed works,

various categorizations of the technologies are presented and discussed. The current challenges and limitations of Assistive Technologies are also briefly outlined.

The review discusses various frameworks and systems that cater to different needs related to ADLs in discrete manners. To use Technology as a Service (TaaS) for independent and assisted living of elderly, it is essential to integrate functionalities of these systems, so that the future of IoT-based technology laden environments can address their various needs related to ADLs which include essential personal activities for Personal Hygiene, Dressing, Eating, Maintaining Continence and Mobility. Till date no such work exists that takes a holistic approach for addressing these challenges. The various challenges associated with the rapidly increasing aging population in the world, as per their diversity and their life stages, creates a looming need for development of a framework or system with multiple functionalities which could include being able to study, learn, track, analyze and respond to elderly behavior to provide personalized assistance and user-centered behavior interventions in the context of ADLs. We believe that this extensive review of assisted living technologies in the context of ADLs which includes discussion, presentation and categorization of the state of art works that currently exist in this field, as well as discussion of the challenges and shortcomings of the same, would serve as the first step for development of such a framework for Ambient Assisted Living of elderly in the future of Smart Homes and Smart Cities.

The proposed work provides a timely discussion about recent advances in this field including directions for future enabling universal access and universal usability of assistive technologies. To the best knowledge of the authors, no similar approach has been done in this field yet. This work for Ambient Assisted Living experiences for elderly people, to be provided by either human caregiver or robotic assistant or both, is to make significant contribution towards the interrelated fields of Human-Computer Interaction, Internet of Things, Human-Robot Interaction, Robotics, Data Science, Machine Learning, Artificial Intelligence, Assistive Systems, Healthcare, Health Information Technologies and their related applications. Future work would involve conducting a comprehensive review of various challenges related to implementation of assistive technologies and discussing remedies and potential measures for addressing the same .